\documentclass{jps-cp}
\usepackage{txfonts} 
\usepackage{bm}

\title{Gluon TMDs and inelastic $J/\psi$ leptoproduction at the EIC}

\author{Umberto \textsc{D'Alesio}$^{1,2}$, Asmita \textsc{Mukherjee}$^{3}$, Francesco \textsc{Murgia}$^{2,*}$, Cristian \textsc{Pisano}$^{1,2}$, and Sangem \textsc{Rajesh}$^{1,2}$}

\inst{$^{1}$Dipartimento di Fisica, Universit\`a di Cagliari, Citt. Universitaria, I-09042 Monserrato (CA), Italy \\
$^{2}$INFN, Sezione di Cagliari, Citt. Universitaria, I-09042 Monserrato (CA), Italy \\
$^{3}$Department of Physics, Indian Institute of Technology Bombay, Powai, Mumbay 400076, India \\
$^{*}$ Speaker}

\email{francesco.murgia@ca.infn.it}

\recdate{February 14, 2022}

\abst{We study the Sivers azimuthal asymmetry for $J/\psi$ production in semi-inclusive deep inelastic scattering, and the still poorly known gluon Sivers function. We concentrate on the $J/\psi$ low transverse momentum region, employing the transverse-momentum dependent generalised parton model (GPM), and its colour gauge invariant extension (CGI-GPM), which includes final-state interactions at leading order. We adopt the nonrelativistic QCD (NRQCD) framework for the quarkonium formation mechanism, and compare our results for the unpolarised cross section and the Sivers asymmetry with available data, respectively from HERA and COMPASS. Finally, we give estimates for the asymmetry in the kinematical regime of the future Electron-Ion Collider (EIC).}

\kword{Gluon Sivers Function, Quarkonium formation mechanism, Electron Ion Collider}

\begin{document}
\maketitle

\section{Introduction}
\label{sec:intro}
Quarkonium production in hadronic collisions is an ideal tool to access the gluon content of the nucleon~\cite{Lansberg:2019adr}.
Collinear gluon parton distributions (PDFs), depending only on the longitudinal fraction of the nucleon momentum carried by the gluon, have already been extensively studied.
Recently, quarkonium production in $pp$ and $\ell p$ collisions has been considered in the context of the transverse momentum dependent (TMD) gluon distributions~\cite{Mulders:2000sh}, particularly for the unpolarised and linearly polarised TMD PDFs and for the gluon Sivers function (GSF). In this contribution we consider $J/\psi$ leptoproduction at small-intermediate transverse momentum as an important tool for extracting complementary information on the GSF. To this end, we adopt a TMD phenomenological approach, the generalised parton model (GPM)~\cite{DAlesio:2004eso}, and the nonrelativistic QCD (NRQCD) effective theory for the quarkonium formation mechanism~\cite{Bodwin:1994jh}. The combined use of these two approaches is challenging, but it can be very helpful in better understanding both the complex three-dimensional structure of hadrons and the details of quarkonium formation.
We will evaluate in this framework both the unpolarised cross section and the Sivers single spin asymmetry for the process $\ell p^\uparrow \to \ell^\prime J/\psi +X$, comparing our results with data from the H1 Collaboration (for the cross section) and with a single data point available from the COMPASS Collaboration for the Sivers asymmetry.
We will then give some estimates for the Sivers asymmetry in kinematical configurations suitable for the Electron Ion Collider (EIC).

\section{Theoretical approach and unpolarised cross section for $\ell p\to \ell^\prime J/\psi + X$}
\label{sec:theo}
In a leading-twist TMD approach there are eight independent gluon TMD  distribution functions (TMD PDFs) for a spin-1/2 hadron~\cite{Mulders:2000sh}. They account for all relevant correlations between the gluon and hadron polarization states and their relative transverse momentum. There is an almost one-to-one formal correspondence with the much better known quark case, considering linearly polarised gluons instead of transversely polarised quarks. The gluon case presents however a more complicated gauge-link colour structure, that is basically at the origin of TMDs. This leads to the existence, at least in principle, of two independent gluon Sivers functions
with different properties and behaviours~\cite{Boer:2015vso}. Their possible interplay in a specific process complicates considerably the phenomenology. However, the two independent GSFs can play different roles in several processes.
Therefore, accurate studies of as much as possible observables involving these distributions, offer a way for disentangling them and gaining first valuable phenomenological information.

In this contribution, we consider the $J/\psi$ leptoproduction process, $\ell p\to \ell^\prime J/\psi + X$, in the $\gamma^*p$ center of mass (cm) reference frame, in which the virtual photon and the target proton move respectively along the $\pm \hat{z}$ axis and the $J/\psi$ is produced
with a small-intermediate transverse momentum.
We denote by $\ell$ and $\ell^\prime$ the 4-momenta of the initial and final leptons, and by $q$, $P$ and $P_h$ those of the virtual photon, the target proton and the produced quarkonium respectively.
We also adopt the usual SIDIS variables: $S=(\ell+P)^2$, $W_{\gamma P}^2=(q+P)^2$, $Q^2=-q^2$, $x_{\rm B}=Q^2/(2P\cdot q)$, $y=(P\cdot q)/(P\cdot\ell)$, $z=(P\cdot P_{h})/(P\cdot q)$. Moreover, we restrict ourselves to the range of $J/\psi$  energy fraction $0.3 < z < 0.9$, covered by H1; the lower cut avoids the resolved-photon region, while the upper one avoids the region of collinear divergences at $z\to 1$~\cite{Mukherjee:2016qxa}.
In order to guarantee the validity of the TMD approach, we will consider $P_T$ values smaller than a few GeV, where $P_T$ is the $J/\psi$ transverse momentum in the $\gamma^*$-$p$ cm frame.
Concerning the unpolarised cross section, we have also checked that moving to larger $P_T$ values, up to around 8 GeV, our TMD estimates smoothly match the corresponding collinear QCD results.
In the TMD GPM plus NRQCD scheme the unpolarised cross section for the process can be written as:
\begin{eqnarray}
\frac{{\rm d}\sigma^{\rm unp}}{{\rm d}Q^2 {\rm d}y {\rm d}^2\bm{P}_T{\rm d}z} &=&
\frac{1}{2S}\,\frac{2}{(4\pi)^4 z}\, \sum_a\,\int\,\frac{{\rm d}x_a}{x_a}\,
{\rm d}^2\bm{k}_{\perp a}\, \delta(\hat{s}+\hat{t} +\hat{u} -M^2+Q^2)\, \nonumber \\
&\times& \sum_n\,\frac{1}{Q^2}\,f_{a/p}(x_a,k_{\perp a})\,L^{\mu\nu}\,H_{\mu\nu}^{a,U}[n]\,
\langle 0 | {\cal O}^{J/\psi}(n) | 0 \rangle\, ,
\label{eq:cross}
\end{eqnarray}
where $M$ is the quarkonium mass; $\bm{k}_{\perp a}$ is the transverse momentum of parton $a$ w.r.t.~the proton direction ($k_{\perp a} = |\bm{k}_{\perp a}|$); $L^{\mu\nu}$ is the unpolarised leptonic tensor; $H_{\mu\nu}^{a,U}$ is the hard-scattering part of the unpolarised hadronic tensor for the virtual photon interacting with parton $a$; the $\langle 0 | {\cal O}^{J/\psi}(n) | 0 \rangle$'s are the NRQCD long-distance matrix elements (LDMEs) for the quarkonium state $n=\, ^1S_0^{(8)}$, $^3S_1^{(1,8)}$, $^3P_J^{(8)}$, $J=1,2,3$ and $(1,8)$ indicates the colour (singlet or octet) state of the $Q\bar Q$ pair evolving into the final colourless quarkonium~\cite{Bodwin:1994jh}. The rest of the notation should be self-explaining.
In the hard-scattering calculations we take into account all partonic contributions at order $\alpha\alpha_S^2$: $\gamma^*+g\to J/\psi+g$, $\gamma^*+q(\bar q)\to J/\psi+q(\bar q)$, including intrinsic charm, and all QED contributions at order $\alpha^3$. Direct $c,\bar c$ fragmentation is negligible in the low-$P_T$ regime considered here. In order to test the dependence of our results from the LDMEs, we consider two sets suitable for the low-intermediate $P_T$ region: the Butenschoen and Kniehl (BK11) set~\cite{Butenschoen:2011yh} and the Sun, Yuan and Yuan (SYY13) one~\cite{Sun:2012vc}, which neglects the colour singlet contribution.
Concerning possible feed-down contributions, we take into account the $\psi(2S)$ one, by adopting the same TMD scheme as for the $J/\psi$ and the LDME set of Sharma and Vitev~\cite{Sharma:2012dy}. Moreover, $\chi_c$ and $b$-quark contributions turn out to be negligible in the low-$P_T$ regime.
In order to proceed, one needs to parametrise the unknown unpolarised quark and gluon TMD PDFs. We consider a simple factorised expression, adopting a Gaussian shape for the tranverse momentum component normalised to unity, with the widths $\langle k_{\perp q}^2\rangle = 0.25$ GeV$^2$, 
 $\langle k_{\perp g}^2\rangle = 1.0$ GeV$^2$ fixed by previous fits to SIDIS and $pp$ data for quark/gluons respectively. For the ordinary, collinear PDFs we consider the CTEQ-L1 set. The factorization scale is taken as $\mu^2  = M^2+Q^2$.
\begin{figure}[tbh]
\centering
\includegraphics[width=0.33\textwidth]{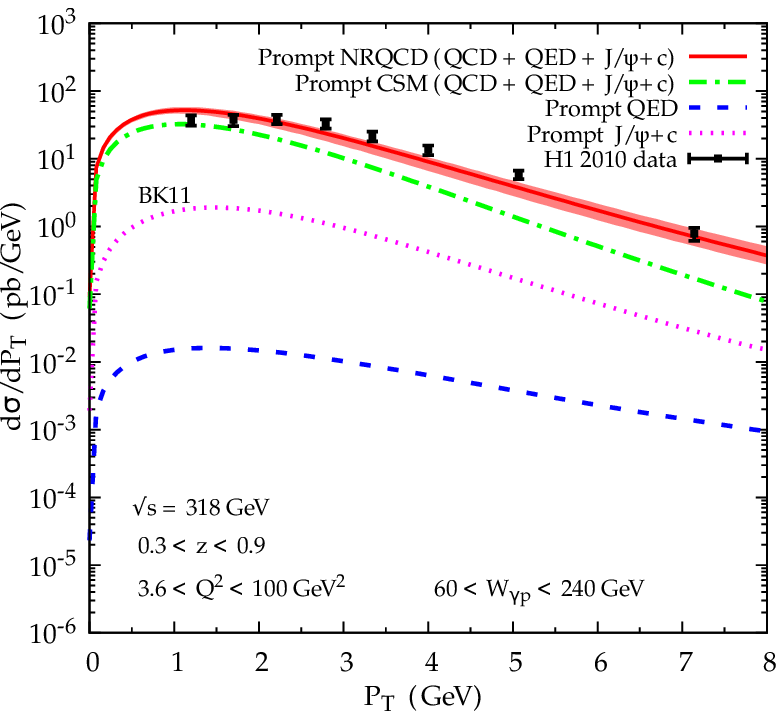}
\hspace{20pt}
\includegraphics[width=0.33\textwidth]{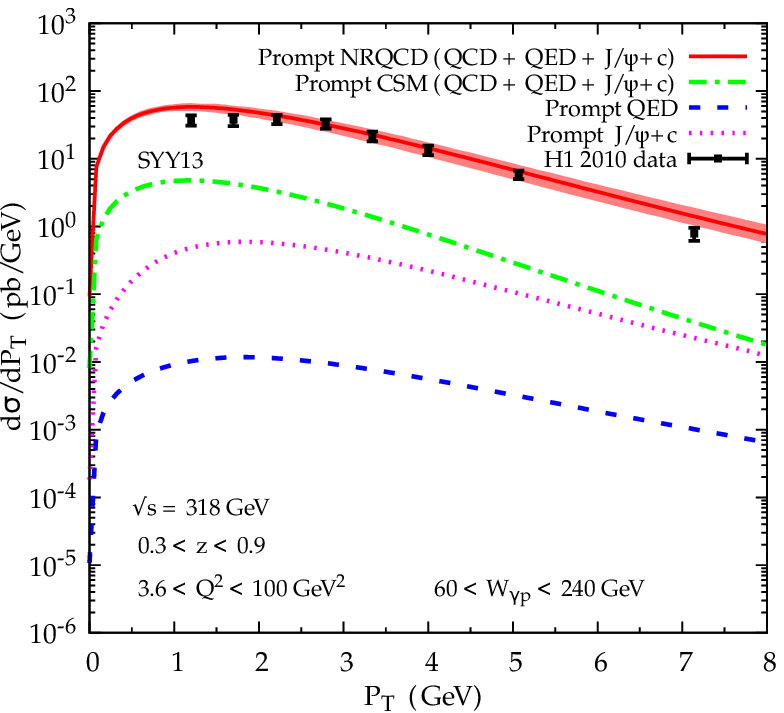}
\caption{Unpolarised cross section for $\ell p\to \ell^\prime J/\psi +X$ in the GPM approach, vs. $P_T$, with the BK11 (left) and SYY13 (right) LDME sets. See the legends for details. Data are from~\cite{H1:2010udv}.}
\label{fig:dsH10}
\end{figure}
In Fig.~\ref{fig:dsH10} we compare our results for the unpolarised cross section, as a function of the $J/\psi$ transverse momentum, with data from the H1 Collaboration at HERA~\cite{H1:2010udv}. We show separately all the partial contributions mentioned above, and the total result, as explained in the legend of the plots, evaluated at the factorization scale $\mu$. The shaded bands on the total result show the uncertainty obtained by varying the scale in the range $[\mu/2\div 2\mu]$. The kinematical configuration is also given in the plots.
These results show that our approach gives estimates in reasonable agreement with data in the low-intermediate $P_T$ region of interest for the study of the Sivers asymmetry and the Sivers gluon function.
Analogous results, and a similar agreement with H1 data, can be obtained for the cross section dependence versus $Q^2$, $W_{\gamma P}$, $z$, in the range $1<P_T<10$ GeV.

\section{Sivers asymmetry and the gluon Sivers function in $\ell p^\uparrow \to \ell^\prime J/\psi + X$}
\label{sec:siv}
The Sivers azimuthal asymmetry for the process $\ell p^\uparrow\to \ell^\prime J/\psi + X$ is defined, in the $\gamma^*$-$p$ cm frame, as follows:
\begin{equation}
A_N^{\sin(\phi_h-\phi_S)} =
2\,\frac{\int {\rm d}\phi_h {\rm d}\phi_S
\sin(\phi_h-\phi_S) ({\rm d}\sigma^\uparrow-
{\rm d}\sigma^\downarrow)}
{\int {\rm d}\phi_h {\rm d}\phi_S 
({\rm d}\sigma^\uparrow+
{\rm d}\sigma^\downarrow)}\,=\,
2\,\frac{\int {\rm d}\phi_h {\rm d}\phi_S
\sin(\phi_h-\phi_S)\,{\rm d}\Delta\sigma(\phi_h,\phi_S)}
{\int {\rm d}\phi_h {\rm d}\phi_S\,
2 {\rm d}\sigma^{\rm unp}}\,,
\label{eq:ansiv}
\end{equation}
where $\phi_h$ and $\phi_S$ are the azimuthal angles, with respect to the leptonic plane determined by $\ell$ and $\ell^\prime$, respectively of the transverse component of the $J/\psi$ momentum, $\bm{P}_T$, and the target proton polarisation vector $\bm{S}$; ${\rm d}\sigma^{\uparrow,\downarrow}$ stay for the polarised differential cross section, defined in analogy to Eq.~(\ref{eq:cross}), along the up and down polarisation directions specified by $\phi_S$; ${\rm d}\sigma^{\rm unp}$ is defined in Eq.~(\ref{eq:cross}).
In the generalised parton model there is only one, hypothetically universal, 
gluon Sivers function. Therefore, the numerator of the Sivers asymmetry, ${\rm d}\Delta\sigma$, can be written as follows:
\begin{eqnarray}
{\rm d}\Delta\sigma^{\rm GPM} &=&\frac{1}{2S}\frac{2}{(4\pi)^4 z}\sum_{a}\int \frac{{\rm d} x_a}{x_a}\, {\rm d}^2{\bm k}_{\perp a}\,\delta\left(\hat{s}+\hat{t}+\hat{u}-M^2+Q^2\right)\left(-2 \, \frac{k_{\perp a}}{M_p}\right)\,\sin(\phi_S-\phi_a)\nonumber\\
 &&{}\times\sum_n\frac{1}{Q^4} \, f_{1T}^{\perp a} (x_a,k_{\perp a})\, L^{\mu\nu}
H^{a,U}_{\mu\nu}[n]\, \langle 0\mid \mathcal{O}^{J/\psi}(n)\mid 0\rangle\, ,
\label{eq:Ds-gpm}
\end{eqnarray}
where $a$ can be a quark or a gluon, $M_p$ is the proton mass, $\phi_a$ is the azimuthal angle of parton $a$ and $f_{1T}^{\perp a}$ is the corresponding Sivers function. Notice that in the GPM approach the hard scattering part is the same as for the unpolarised case.
In the colour gauge invariant GPM (CGI-GPM) things get more complicated: because of initial- (ISIs) and final- (FSIs) state interactions, accounted for through appropriate colour gauge links, TMD PDFs become in general process dependent~\cite{DAlesio:2020eqo}; moreover, in the gluon case, there are in principle two independent gluon Sivers functions, the so-called $f$-type and $d$-type ones~\cite{Boer:2015vso}. The (perturbatively calculable) process dependent colour factors can be absorbed into modified hard-scattering terms, related to the unpolarised ones as follows:
\begin{equation}
H^{{\rm Inc}(f/d)} = \frac{C_I^{(f/d)} + C_F^{(f/d)}}{C_U}\,H^U\, ,
\label{eq:Hinc}
\end{equation}
where $C_{I,F}^{(f,d)}$ are the modified $f$- and $d$-type colour factors when including ISIs and FSIs respectively, while $C_U$ is the one for the unpolarised case. 
In particular, for $J/\psi$ leptoproduction there are no ISIs, $C_{I}^{(f,d)}=0$; in addition, $C_{F}^{(f,d)}$ vanish for the colour singlet quarkonium state, that is FSIs contribute only for colour octet states; moreover, $C_{F}^{(d)}=0$, therefore only the $f$-type GSF contributes to the process under consideration. This is very important from the phenomenological point of view, since it allows us to get direct information on the $f$-type GSF.
In the CGI-GPM the numerator of the Sivers asymmetry can therefore be written as:
\begin{eqnarray}
{\rm d}\Delta\sigma^{\mathrm{CGI-GPM}} &\!\!\!=& \frac{1}{2S}\frac{2}{(4\pi)^4 z}\int \frac{{\rm d} x_a}{x_a}\, {\rm d}^2{\bm k}_{\perp a}\,\delta\left(\hat{s}+\hat{t}+\hat{u}-M^2+Q^2\right) \left(-2 \, \frac{k_{\perp a}}{M_p}\right)\,\sin(\phi_S-\phi_a)\nonumber\\
&\!\!\!\!\!\!\!\!\!\!\!\!\!\!\!\!\!\!\!\!\!\!\!\!\!\!\!\!\!\!\!\!\!\!\!\!\times&\!\!\!\!\!\!\!\!\!\!\!\!\!\!\!\!\!\!\!\!\!\sum_{n}\frac{1}{Q^4}\,L^{\mu\nu} \Big\{\sum_q f_{1T}^{\perp q} (x_a, k_{\perp a})\,H^{q,\mathrm{Inc}}_{\mu\nu}[n]+ f_{1T}^{\perp(f) g} (x_a, k_{\perp a})\, 
H^{g,\mathrm{Inc}(f)}_{\mu\nu}[n]\Big\}\, \langle 0\mid \mathcal{O}^{J/\psi}(n)\mid 0\rangle\,.
\label{eq:Ds-cgi}
\end{eqnarray}
We parametrise the Sivers function in a simple factorised form, taking the $x$-dependent component proportional to the collinear unpolarised PDF and the $k_\perp$-dependent one in a  Gaussian-like form:
\begin{equation}
\Delta^N f_{a/p^\uparrow}(x_a,k_{\perp a},\mu) =
- 2\frac{k_{\perp a}}{M_p} f_{1T}^{\perp a}(x_a,k_{\perp a},\mu) =
2{\cal N}_a(x_a) f_{a/p}(x_a,\mu)\frac{\sqrt{2e}}{\pi} \sqrt{\frac{1-\rho_a}{\rho_a}} k_{\perp a}
\frac{e^{-k_{\perp a}^2/(\rho_a \langle k_{\perp a}^2\rangle)}}{\langle k_{\perp a}^2\rangle^{3/2}}\, ,
\label{eq:siv}
\end{equation}
where $\rho_a = \langle k_{\perp a}^2\rangle_S/( \langle k_{\perp a}^2\rangle + \langle k_{\perp a}^2\rangle_S)$ and $ \langle k_{\perp a}^2\rangle_S$ is a new parameter modulating the Gaussian shape of the Sivers function.
We present estimates for the maximised Sivers asymmetry in $J/\psi$ leptoproduction for COMPASS and EIC kinematical configurations respectively. To this end, we saturate the positivity bound for the collinear component of the Sivers function by taking ${\cal N}_q(x) = {\cal N}_g^{(f,d)}(x) = +1$, and maximise the transverse component contribution by taking $\rho_a=2/3$ for both quarks and gluons.

\begin{figure}[bth]
\begin{center}
\hspace{-35pt}
\includegraphics[width=0.40\textwidth]{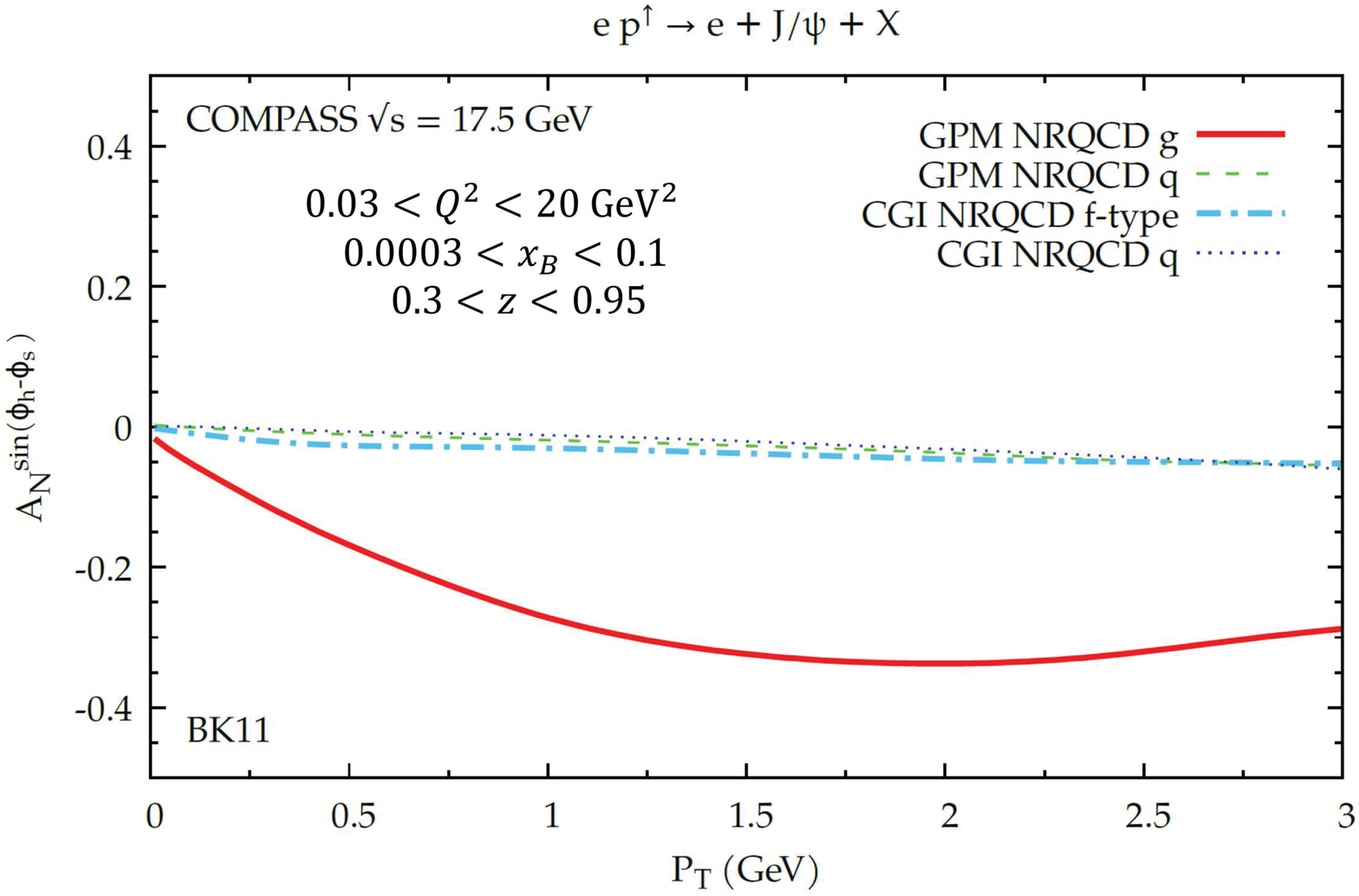}
\hspace{35pt}
\includegraphics[width=0.40\textwidth]{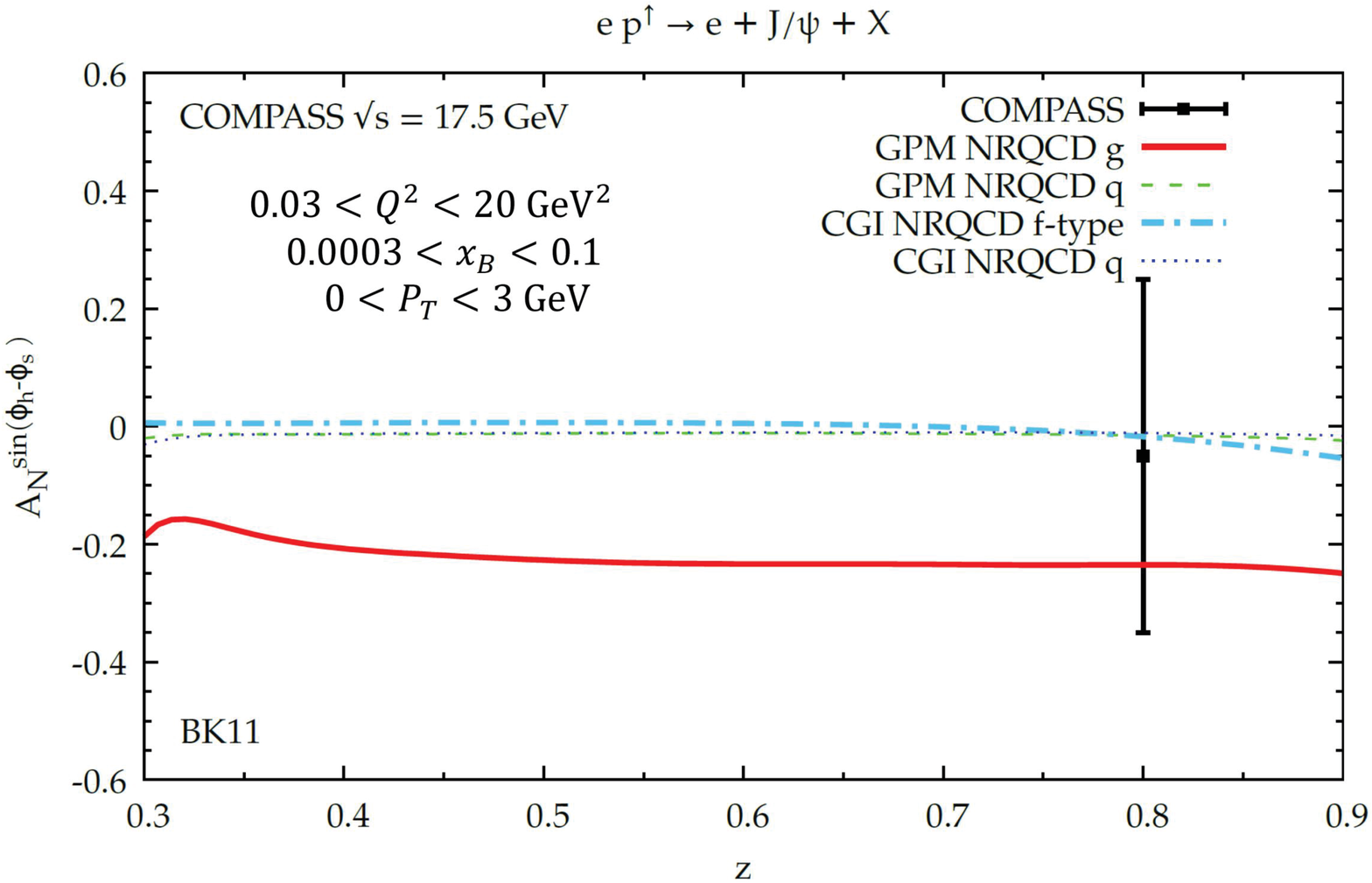}
\end{center}
\caption{Maximised Sivers asymmetry for $\ell p^\uparrow \to \ell^\prime J/\psi + X$, with the BK11 LDME set, vs.~$P_T$ (left) and $z$ (right), for COMPASS kinematics. See the legends for details. The data point is from~\cite{Matousek:2016xbl}.}
\label{fig:siv-comp}
\end{figure}

In Fig.~\ref{fig:siv-comp} we show the maximised Sivers asymmetry, both in the GPM and CGI-GPM cases and adopting the BK11 LDME set, for the COMPASS kinematical configuration, as a function of $P_T$ (left) and $z$ (right), where the only COMPASS data point available, with large errors, is also shown~\cite{Matousek:2016xbl}.

\begin{figure}[bth]
\begin{center}
\hspace{-50pt}
\includegraphics[width=0.40\textwidth]{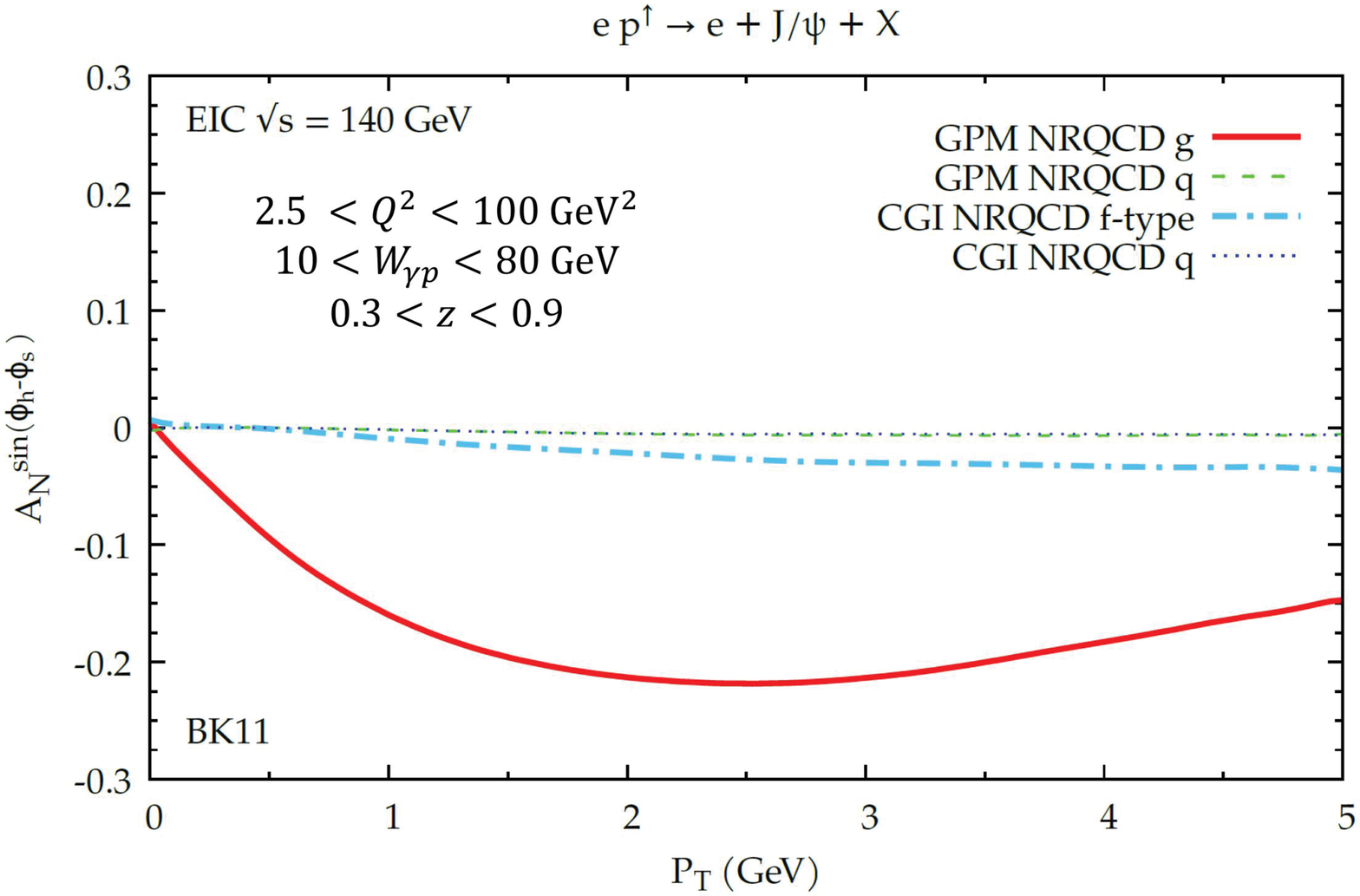}
\hspace{25pt}
\includegraphics[width=0.40\textwidth]{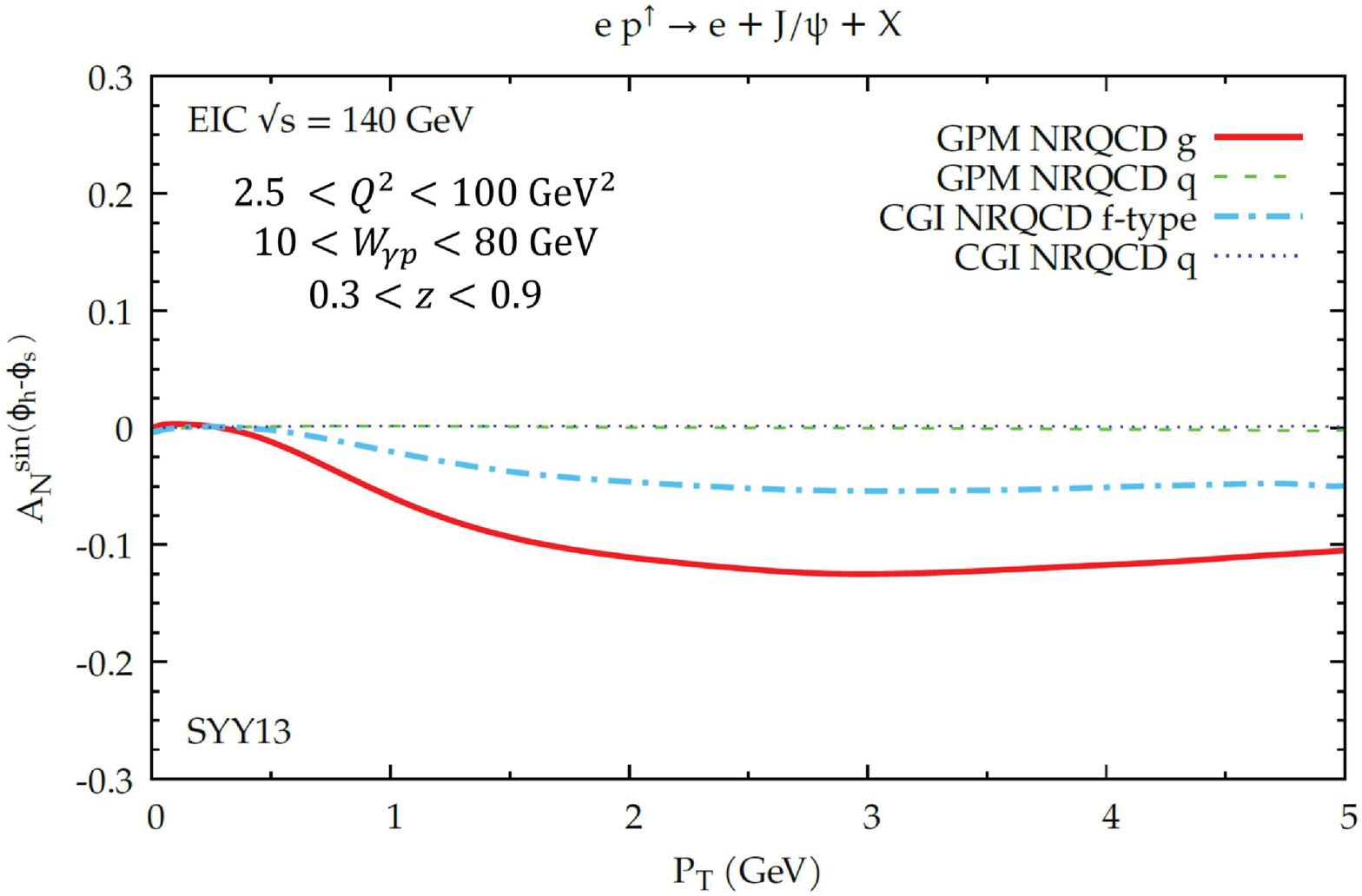}
\end{center}
\caption{Maximised Sivers asymmetry for $\ell p^\uparrow \to \ell^\prime J/\psi + X$, vs. $P_T$, with the BK11 (left) and SYY13 (right) LDME sets, for EIC kinematics at $\sqrt{s}=140$~GeV. See the legends for details.}
\label{fig:siv-eic140}
\end{figure}

In Fig.~\ref{fig:siv-eic140} we present estimates for the maximised Sivers asymmetry, both for the GPM and CGI-GPM cases for a typical EIC kinematical configuration, as a function of $P_T$, adopting the BK11 (left) and SYY13 (right) LDME sets.
Details of the kinematics are given in the legends.
All the plots show the same general trend: the gluon contribution in the GPM can be potentially large in size and negative, with an opposite sign w.r.t.~that chosen for the $f$-type GSF.
All other contributions, that is the quark ones (both in the GPM and CGI-GPM cases) and the gluon $f$-type one in the CGI-GPM approach are either very small or even negligible. However, it is interesting to notice that with the SYY13 LDME set the gluon contribution in the CGI-GPM case can potentially reach a size of about 0.05 at intermediate $P_T$ values, still with a negative sign if the $f$-type GSF is taken positive. 

\section{Conclusions}
In this contribution we have considered $J/\psi$ electroproduction as a tool for studying gluon TMDs and the Sivers asymmetry. To this aim we have adopted a (colour gauge invariant) generalised parton model complemented with the NRQCD effective theory for the quarkonium formation  mechanism.  
We have investigated the dependence of our results on the LDME set adopted, by considering two available sets covering the $P_T$ region of interest for the TMD approach. Reasonable agreement with H1 data for the unpolarised cross section in the small-intermediate $P_T$ range has been found. Concerning the Sivers asymmetry in the CGI-GPM, we showed that only colour-octet FSIs and the $f$-type GSF are at work in this process, which can therefore be very useful phenomenologically. Lastly, we have presented some preliminary results for the maximised Sivers asymmetry in typical COMPASS and EIC kinematical configurations.
A more detailed discussion and an extended phenomenological analysis are in progress and will be presented elsewhere~\cite{paper}.

\end{document}